\documentclass[prb,aps,twocolumn,groupedaddress,floats,showpacs,final,superscriptaddress]{revtex4-1}
\usepackage[latin3]{inputenc}
\usepackage[makeroom]{cancel}
\usepackage{graphicx}
\usepackage{amsmath}
\usepackage{amsfonts}
\usepackage{amssymb}
\usepackage{chngcntr}
\usepackage{color}
\usepackage[left=2cm,right=2cm,top=2cm,bottom=2cm]{geometry}

\usepackage{graphicx}
\usepackage{dcolumn}
\usepackage{bm}
\usepackage{simplewick}
\usepackage{array}
\usepackage{appendix}

\RequirePackage[
   hyperindex,colorlinks,bookmarksnumbered,
   plainpages=true,pdfstartview=FitH]{hyperref}
\hypersetup{linkcolor=blue,urlcolor=blue,citecolor=blue}
\usepackage{hyperref}

\definecolor{purple}{rgb}{0.5,0,0.6}

\usepackage{ulem}

\begin{document}

\date{\today}
\title{Pumping and cooling of nanomechanical vibrations generated by Cooper pair exchange}
\author{Anton V. Parafilo}
\email{aparafil@ibs.re.kr}
\affiliation{Center for Theoretical Physics of Complex Systems, Institute for Basic Science, Expo-ro, 55, Yuseong-gu, Daejeon 34126, Republic of Korea}

\author{Leonid Y. Gorelik}
\affiliation{Department of 
Physics, Chalmers University of
Technology, SE-412 96 G{\" o}teborg, Sweden}

\author{Hee Chul Park}
\email{hc2725@gmail.com}
\affiliation{Center for Theoretical Physics of Complex Systems, Institute for Basic Science, Expo-ro, 55, Yuseong-gu, Daejeon 34126, Republic of Korea}

\author{Robert I. Shekhter}
\affiliation{Department of Physics, University of Gothenburg, SE-412
96 G{\" o}teborg, Sweden}

\date{\today}

\begin{abstract}
We consider a nanoelectromechanical system composed of a carbon nanotube suspended between two normal leads and coupled to a
superconducting scanning tunneling microscope (STM) tip via vacuum tunnel barrier. Treating the nanotube as a single-level quantum dot, it is shown that an
applied voltage between the superconducting STM tip and normal leads gives rise to a pumping or a cooling of the mechanical subsystem
depending on the direction of the electronic flow.
It is also demonstrated that the transition between these two regimes is controlled
by the strength of the tunnel coupling between the nanotube and superconducting STM tip and the relative position of the electronic level. Such phenomena are realized due to a specific electromechanical coupling that is fully governed by the
quantum dynamics of the
Cooper pairs. The amplitude of the self-sustained oscillations in the pumping regime is analyzed numerically, and the effective temperature of the mechanical subsystem in the cooling regime is obtained.
\end{abstract}
\maketitle

\section{ Introduction} Nanoelectromechanical systems (NEMS) provide a
promising platform for investigations of the quantum mechanical
interplay between mechanical and electronic subsystems 
\cite{cleland},\cite{ekinci}. The generation of self-driven mechanical
oscillations by a dc electronic flow \cite{shuttle},\cite{blanter}
is a bright exhibition of such interplay. The study of self-driven phenomena is itself an interesting problem from a fundamental point of view since such phenomena are promising for mass and force
sensing, while the underlying physical processes have potential
applications for mechanical cooling \cite{cooling1}-\cite{coolingexp}.
Self-driven mechanical oscillations were first observed in a carbon
nanotube (CNT)-based transistor \cite{huttel1}, and their transport
signatures were later verified in further studies
\cite{huttel2},\cite{huttel3}. Recently, an experimental observation of
self-driven oscillations in a nanomechanical suspended
CNT-based resonator in the Coulomb blockade regime were reported 
\cite{exp},\cite{self}.

The crucial point for NEMS performance is the nature of
the coupling between the nanomechanical and electronic subsystems.
Usually, coupling is associated with the localization of electronic charge
or spin on the movable parts \cite{parafiloreview},\cite{antonreview}, while at the same time it can also be associated with the so-called {\it covalent coupling} \cite{ziman} (or covalent bonding) based on the sharing of electrons between atoms and molecules that is well known in chemistry.
Superconducting (SC) elements incorporated into NEMS, the ground state of which may be considered as a macroscopic
"molecule", open the possibility to consider this type of coupling as an origin for electromechanical performance. Namely, a
SC lead located near a moving quantum dot (QD) can affect its electronic
state due to the tunneling exchange of Cooper pairs (SC proximity effect). Moreover, if
the tunneling amplitude depends on the distance between the QD and
the SC lead, such exchange also provides a connection between the
electronic and mechanical degrees of freedom. An additional injection
of electrons from a biased normal metal electrode into the QD
generates peculiar dynamics of the Cooper pairs on it. 
As a consequence, the interplay between coherent two-electron (Cooper
pair) and incoherent single-electron tunneling into/out of the
movable part of the NEMS 
results in both pumping and cooling effects.

The paper is organized as follows. Section \ref{model} is devoted to the formulation of the model describing a hybrid superconducting--normal metal nanoelectromechanical device. The reduced density matrix technique and a Wigner function description of the mechanical subsystem are discussed in Sec. \ref{3}.  Details about the mechanical subsystem in the two regimes of cooling and pumping are provided in Sec. \ref{cooling}, and a summary and discussions are given in Sec. \ref{conclusion}.

\begin{figure}
\centering \includegraphics[width=1\columnwidth]{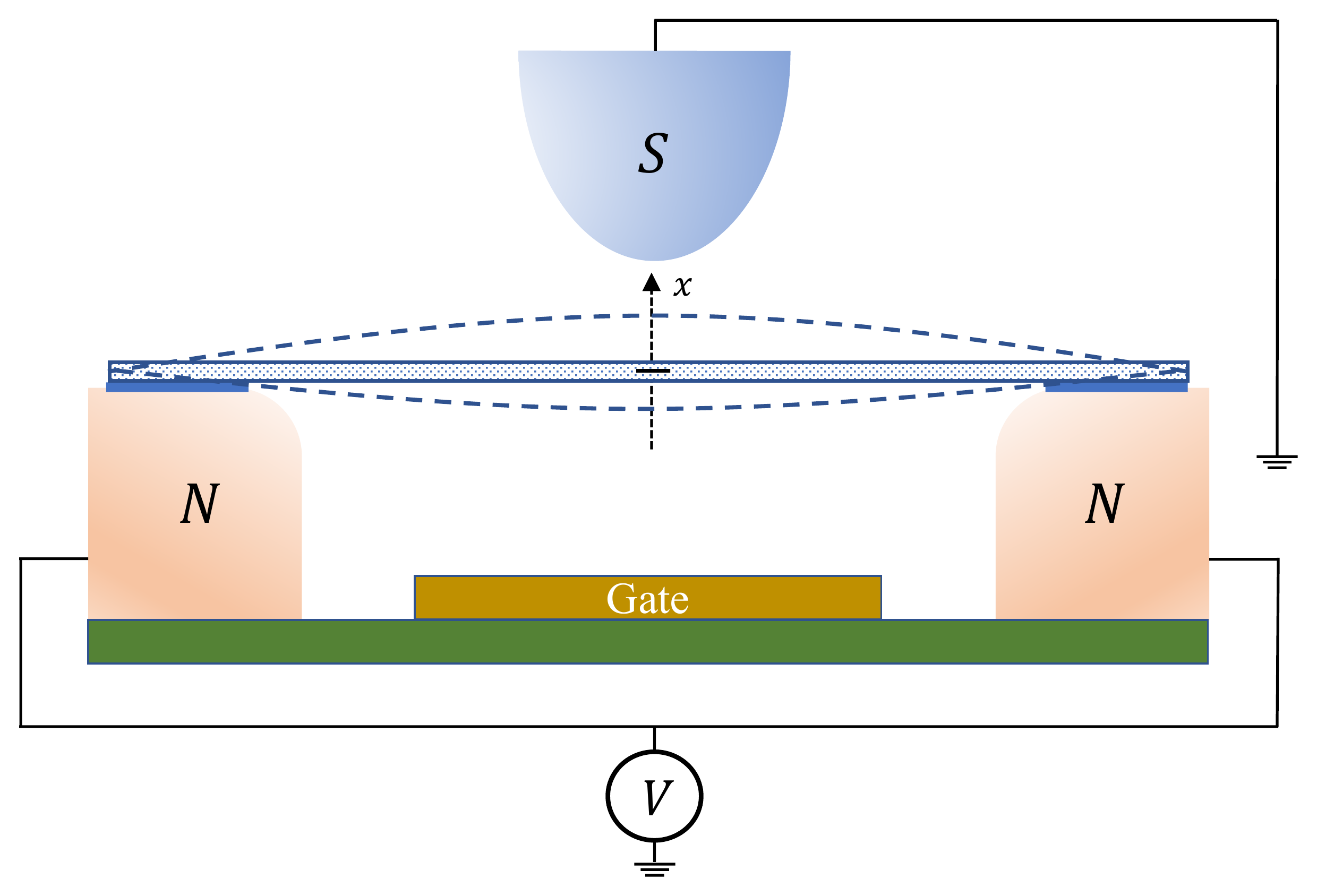} \caption{Scheme of the superconducting--normal metal
hybrid nanoelectromechanical device. A suspended single-wall carbon nanotube (CNT) is placed between two
normal metal leads equally biased with voltage $V$. Under the considered set of parameters, the CNT is treated as a
two-fold degenerate single-level quantum dot (QD). A grounded superconducting lead (STM tip) placed near the CNT-QD
induces the proximity effect in the QD and thus provides for electromechanical  coupling via the CNT's bending-dependent tunnel amplitude (see the main text).}
\label{Fig1}
\end{figure}

\section{Model}\label{model}
A sketch of the NEMS under investigation is presented
in Fig.~\ref{Fig1}. We consider a nanoelectromechanical device
consisting of a metallic single-wall CNT suspended
between two equally voltage-biased normal electrodes and coupled to a grounded superconducting scanning tunneling microscope (STM) tip via vacuum tunnel barrier. We assume
that the CNT is short enough to operate in the regime where the
electronic mean-level spacing is greater than the temperature $k_BT$
and bias voltage $eV$.
The CNT bending, which crucially affects the tunnel coupling with the STM tip, is 
described by the fundamental flexural mode. These assumptions allow us to treat the CNT as a movable single-level
QD whose position relative to the SC tip
is free to vary.

The Hamiltonian of the model reads as follows:
\begin{eqnarray}\label{hamiltonian}
H=H_{N}+H_S+H_{QD}+H_{tun}.
\end{eqnarray}
The first two terms in Eq.~(\ref{hamiltonian}) describe all
the electrodes in the system: the two normal metal leads that are equally biased by voltage $V$, and the SC STM tip characterized by an order parameter $\Delta$:
\begin{eqnarray}\label{hamleads}
&&H_N=\sum_{kj\sigma} (\varepsilon_k-eV)c^{\dag}_{kj\sigma}c_{kj\sigma},\\
&&H_{S}=\sum_{k \sigma}\left\{\xi_{k} a^{\dag}_{k\sigma }a_{k\sigma
}-\Delta(a^{\dag}_{k\uparrow }a^{\dag}_{-k\downarrow
}+H.c.)\right\}.
\end{eqnarray}
Here, $c_{kj\sigma}$ ($c_{kj\sigma}^{\dag}$) and $a_{k\sigma}$
($a_{k\sigma}^{\dag}$)  are annihilation (creation) operators of the electrons with spin ($\sigma=\uparrow,\downarrow$) in the normal ($j=L$ and $j=R$ stand for the left and right electrodes, respectively) and
SC leads with energies $\varepsilon_k$ and $\xi_{kj}$,
correspondingly.

The Hamiltonian of the single-level vibrating CNT-QD reads as follows:
\begin{eqnarray}\label{hamQD}
H_{QD}=\sum_{\sigma}&&\varepsilon_0
d^{\dag}_{\sigma}d_{\sigma}+\frac{\hbar\omega_0}{2}(\hat p^2 + \hat
x^2) .
\end{eqnarray}
The first term in Eq.~(\ref{hamQD}) describes the quantum dynamics of the electronic
degree of freedom: $\varepsilon_0$ is the QD electron energy level, and $d_{\sigma}$, $d_{\sigma}^{\dag}$ are annihilation and creation
operators of the electrons in the QD.
The second term in Eq.~(\ref{hamQD}) characterizes the linear
dynamics of the fundamental flexural mode, which we treat as a mechanical
oscillator with frequency $\omega_0$. Dimensionless operators
$\hat x=\hat X/x_0$, $\hat p=x_0 \hat P/\hbar$ are canonically
conjugated displacement and momentum, where $x_0=\sqrt{\hbar/m\omega_0}$
is the amplitude of the zero-point oscillations with $m$ the effective
mass of the CNT.

The last term in Eq.~(\ref{hamiltonian}),
\begin{eqnarray}\label{tun}
H_{tun}=\sum_{k\sigma}e^{ \zeta\hat{x}/2}\left(t_k^s a^{\dag}_{k\sigma}d_{\sigma}+(t_k^s)^{\ast} d^{\dag}_{\sigma}a_{k\sigma}\right)\nonumber\\
+\sum_{kj\sigma}\left(t_k^n
c^{\dag}_{kj\sigma}d_{\sigma}+(t_k^n)^{\ast}
d^{\dag}_{\sigma}c_{kj\sigma}\right),
\end{eqnarray}
describes the tunneling processes between the CNT and i) the STM tip with a
deflection-dependent hopping amplitude $t_k^s \exp(\zeta\hat x)$
($\zeta\sim x_0/l$ and
 $l \simeq 10^{-10}\,{\rm m}$ is the tunneling length of the barrier), see the first line in Eq.~(\ref{tun}), and ii) the normal leads with a tunnel amplitude $t_k^n$, see the second line in Eq.~(\ref{tun}). Note that the deflection-dependent tunneling amplitude causes the electromechanical coupling in the above model.

\section{Reduced density matrix and Wigner function description}\label{3} 
To analyze the stationary state of the mechanical subsystem in the
SC--normal metal hybrid junction described by the Hamiltonian in Eq.~(\ref{hamiltonian}),
we use the {\it reduced density matrix technique} \cite{novotny},\cite{fedorets}. First
we obtain the quantum master equation for the total density matrix
$\hat{\varrho}$ by considering the tunneling Hamiltonian in Eq.~(\ref{tun}) as a perturbation. Then we look for the solution of this equation in
the form
$\hat{\varrho}=\hat{\rho}_{S}\bigotimes\hat{\rho}_{N}\bigotimes\hat{\rho}$, where $\hat{\rho}_{S}$ and  $\hat{\rho}_{N}$  are density matrices
describing the thermodynamic state of the SC and normal
electrodes, while density matrix $\hat{\rho}$ describes the
mechanical and electronic states of the movable single-level QD. After substituting this {\it anzats} into the obtained master equation and tracing out the electronic degrees of freedom in both normal and SC leads, we get the equation for the reduced density
matrix $\hat{\rho}$, which reads as follows (in $\hbar=k_B=1$ units):
\begin{eqnarray}\label{mast}
\dot {\hat\rho} =&& -i[H_{QD},\hat\rho]+i\Delta_{d}[e^{\zeta
\hat x}(d^{\dag}_{\uparrow}d^{\dag}_{\downarrow}+h.c.),\hat\rho]-\mathcal{L}_{e}[\hat\rho].
\end{eqnarray}
Here, $\Delta_{d}=2\pi \nu_0^{s} |t_k^s|^2$ is the strength of the
intra-QD electron pairing generated by the proximity effect with the SC STM tip,
$\nu_0^{s}$ is the normal density of states of electrons in the STM,
 and $\mathcal{L}_e[\hat\rho]$ is a Lindbladian superoperator describing the incoherent electronic exchange between the CNT and normal biased electrodes, which in the
 high-voltage regime $eV\gg 2|\varepsilon_0|, \omega_0$ considered in this
 paper takes the form:
\begin{eqnarray}\label{lindbladian}
\mathcal{L}_{e}[\hat\rho]=\frac{\Gamma}{2}\left\{\begin{array}{c}
 \sum_{\sigma}(\{d_{\sigma}d_{\sigma}^{\dag},\hat\rho\}-2d^{\dag}_{\sigma}\hat\rho d_{\sigma}),V>0,\\
\sum_{\sigma}(\{d^{\dag}_{\sigma}d_{\sigma},\hat\rho\}-2d_{\sigma}\hat\rho
d^{\dag}_{\sigma}),V<0,
\end{array}\right.
\end{eqnarray}
where $\Gamma=2\pi \nu^{n}_0 |t_k^n|^2$ is the QD energy level
width and $\nu_0^{s}$ is the electron density of states in the normal
electrodes. The direction of the electronic flow (sign of the bias voltage) is characterized by the parameter $\kappa={\rm sign}(V)$ in what follows. 

The quantum master equation [Eq.~(\ref{mast})] is justified in the so-called {\it
deep sub-gap regime} \cite{nqds},\cite{atomlasing}, which assumes that all energies are smaller
than the SC gap ($eV, T, \varepsilon_0\ll \Delta$) and disregards all scattering
processes above the SC gap. Interestingly, the nanoelectromechanical coupling, see the second term in Eq.~(\ref{mast}), only exists as long as there is a tunnel connection with the SC lead, $\Delta_d\neq 0$.

The reduced density matrix $\hat\rho$ acts in a Hilbert space, being a tensor product
of a finite Fock space of the two-fold
degenerate single-electron level quantum state and ${\rm L}^{2}(x)$ space describing
the mechanical degree of freedom. This Hilbert space is treated as a
direct sum of four ${\rm L}^{2}(x)$ sub-spaces:
$|0,x\rangle=|0\rangle\bigotimes|x\rangle$,
$|\sigma,x\rangle=d^{\dag}_{\sigma}|0\rangle\bigotimes|x\rangle$
($\sigma=\uparrow,\downarrow$), and $|2,x\rangle=d^{\dag}_{\uparrow}d^{\dag}_{\downarrow}|0\rangle\bigotimes|x\rangle$.

\begin{figure}
\centering
\includegraphics[width=0.9\columnwidth]{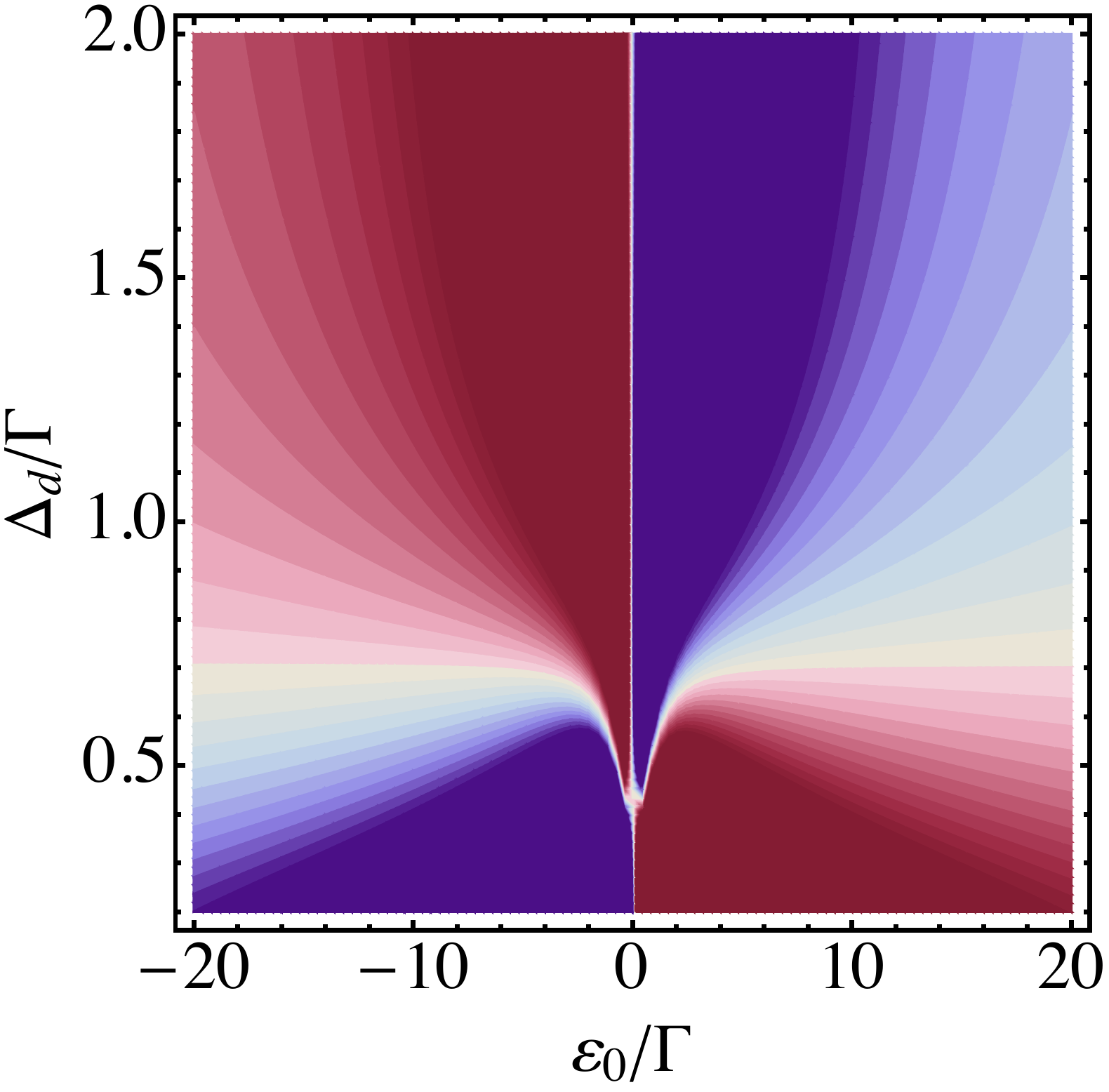} \caption{Effective inverse temperature $\beta_{eff}(0)=(T_{eff}/\omega_0)^{-1}$, see Eq.~(\ref{beta}), as a function of the CNT-QD's energy level position $\varepsilon_0$ and normalized tunnel coupling with the superconducting STM tip $\Delta_d/\Gamma$. The blue color scheme indicates the regime of cooling (i.e., a ground state solution with the amplitude of mechanical oscillations $A=0$ is stable), while the red color scheme indicates the regime of self-sustained oscillations (the solution with $A=0$ is unstable). White regions separate the regimes. The diagram is plotted at fixed bias voltage polarity $\kappa=+1$.}
\label{Fig2}
\end{figure}

Below we investigate the stationary solutions of the the master
equation, Eq.~(\ref{mast}). In order to do this it is convenient to use the
Wigner function representation $w_{i,i'}(x ,p)= \int dy
\rho^{i,i'}(x-\frac{y}{2},x+\frac{y}{2})e^{iyp}$ for the matrix
elements of the density operator $\rho^{i,i'}(x,x')=\langle
i,x\vert\hat\rho\vert i',x'\rangle$ ($i,i'=0, \uparrow, \downarrow,2$). In
general, Eq.~(\ref{mast}) generates a system of linear
partial differential equations for the 16 real functions of $x$ and
$p$. However, the number of relevant functions may be reduced by taking into account that
$\rho^{\sigma,0}(x,x')=\rho^{\sigma,2}(x,x')=\rho^{\uparrow,\downarrow}(x,x')=0$. We utilize this condition because of the superselection rule that forbids a quantum superposition of states with different parity. The time-reversal symmetry additionally results in
$\rho^{\uparrow,\uparrow}(x,x')=\rho^{\downarrow,\downarrow}(x,x')$.

As a consequence, the system of the remaining equations can be presented as a linear combination of five functions: $W_{\Sigma}(x ,p)= \sum_{i}w_{i,i}(x, p)$, $W_{0}(x,p)=
w_{0,0}(x, p)+w_{2,2}(x, p)$, $W_{1}(x, p)=2 \mathrm{Re} [w_{2,0}(x, p)]$,
$W_{2}(x, p)= - 2 \mathrm{Im} [w_{2,0}(x, p)]$, and $W_{3}(x, p)=
w_{2,2}(x, p)-w_{0,0}(x, p)$, and reads as follows:
\begin{eqnarray}
&&\hat{L}W_{\Sigma} = \zeta \Delta_{d}(x)\frac{\partial W_1}{\partial p},\label{eq1}
\\
&&(\hat{L}-2\Gamma)W_{0} = \zeta \Delta_{d}(x)\frac{\partial W_1}{\partial p}-\Gamma\left(\kappa W_3+W_{\Sigma}\right),\label{eq2}
\\
&&(\hat{L}+\hat M)\vert {\bf W}\rangle=
\left(\begin{array}{c}
\Delta_d(x)\zeta \frac{\partial W_0}{\partial p}\label{eq3}\\
0
\\
-\kappa\Gamma W_{\Sigma}
\end{array}\right).
\end{eqnarray}
Here, $\hat L$ is a differential operator that reads as
\begin{equation}\label{}
\hat{L}=\omega_0\left(x\frac{\partial}{\partial p}+p\frac{\partial}{\partial x}\right),
\end{equation}
and we define the vector $\vert {\bf W}\rangle=(W_1, W_2, W_3)^T$ and use a shorthand notation $\Delta_d(x)=\Delta_d \exp(\zeta x)$.
In Eq.~(\ref{eq3}) we introduce a matrix $\hat M=- \Gamma \hat 1 -i 2\varepsilon_0 \hat\lambda_2+i 2\Delta_d (x)\hat \lambda_7$, where $\hat 1$ is the unit matrix and $\hat \lambda_2, \hat \lambda_7$ are the Gell-Mann matrices.

To 
analyze the system of Eqs.~(\ref{eq1})--(\ref{eq3}), we use the perturbation theory over small mechanical frequency $\omega_0$ and present the vector $|{\bf W}\rangle$ approximately as $|{\bf W}\rangle \approx |{\bf W}^{(0)}\rangle+(\omega_0/\Gamma)|{\bf W}^{(1)}\rangle$.
Thus, in the first order of perturbation theory, one can combine Eqs.~(\ref{eq1})--(\ref{eq3}) into a form of the Fokker--Planck equation for $W_{\Sigma}(x, p)$:
\begin{widetext}
\begin{eqnarray}\label{FPequation}
&&\left[\hat L - d(x)\frac{\partial}{\partial p}-\zeta \Delta_{d}(x) \frac{\partial}{\partial p} \left(f(x)\frac{\partial}{\partial p}+\gamma(x)+ R(x)\hat L\right)\right]W_{\Sigma}(x, p)=0,
\end{eqnarray}
\end{widetext}
where $ d(x)=-\kappa\zeta\Gamma\Delta_d(x)\langle e_1|\hat M^{-1}|e_3\rangle$ is the equilibrium displacement of the CNT induced by the SC proximity effect, and
\begin{eqnarray}\label{coefficients}
&&f(x)=\zeta\langle e_1|\hat M^{-1}|e_1\rangle\Delta_d(x)C_0,\\
&&\gamma(x)=\kappa\Gamma  \langle e_1| \hat M^{-1}\hat L \hat M^{-1}|e_3\rangle,\\
&&R(x)=\kappa \Gamma\langle e_1| \hat M^{-2}|e_3\rangle .\label{coefficients1}
\end{eqnarray}
Here, we denote the vectors $|e_1\rangle=(1,0,0)^{T}$, $|e_3\rangle=(0,0,1)^{T}$, and consider the fact that $\hat M |{\bf W}^{(0)}\rangle=-\kappa\Gamma |e_3\rangle W_{\Sigma}$ and $W^{(0)}_0=C_0 W_{\Sigma}$.

First, we eliminate
the shift $\propto d(x)$ by redefining the CNT's displacement coordinate: $x\rightarrow x+x_{eq}$, where $x_{eq}$ can be found by solving the equation $\omega_0 x_{eq}= \zeta d (x+x_{eq})$. Since the equilibrium displacement $x_{eq}\approx - \zeta (\kappa\varepsilon_0/\omega_0)[\Delta_d^2/(\varepsilon_0^2+\Delta_d^2+\Gamma^2/4)]$ appears as a consequence of the SC proximity effect, it disappears if the coupling with the SC lead vanishes, and it additionally depends on the sign of both the CNT-QD energy level $\varepsilon_0$ and bias voltage polarity $\kappa$. Since $x_{eq}\sim \zeta$, it gives only a small renormalization of $\Delta_d(x)$, which we ignore in what follows.

Second, developing the perturbation theory over small parameter $\zeta \sim 10^{-2}\div 10^{-3}$, one can find the solution of Eq.~(\ref{FPequation}) in the form:
\begin{eqnarray}\label{solution}
W_{\Sigma}\approx W_{\Sigma}^{(0)}(x^2+p^2)+O(\zeta^2 ).
\end{eqnarray}
Solving Eq.~(\ref{FPequation}) consistently in all orders of perturbation theory and using for convenience an action-angle representation ($x = A \cos\varphi$, $p=A\sin\varphi$), one can simplify Eq.~(\ref{FPequation}) as follows,
\begin{eqnarray}\label{difeq}
\left\{\tilde\gamma(A)A+\tilde D(A)\frac{d}{dA}\right\}W_{\Sigma}^{(0)}(A)=0.
\end{eqnarray}
Here, the notations $\tilde \gamma (A)=\int_{-\pi}^{\pi}d\varphi \sin^2\varphi \,\Delta_d(A\cos\varphi)\gamma(A\cos\varphi)$, $\tilde D (A)=\zeta^2\int_{-\pi}^{\pi}d\varphi \sin^2\varphi \,\Delta_d(A\cos\varphi)D(A\cos\varphi)$, and  $D(x)=f(x)+\omega_0 x_{eq}R(x)$ are introduced. For simplicity we also redefine the dimensionless amplitude taking $A \zeta \rightarrow A$. The solution of Eq.~(\ref{difeq}) is easy to find and reads as follows,
\begin{eqnarray}\label{stationsolution}
W_{\Sigma}^{(0)}(A)=\mathcal{Z}^{-1} \exp\left[-\int_0^A dA' A'\frac{\tilde\gamma(A') }{\tilde D(A')}\right],\nonumber\\
\end{eqnarray}
where $\mathcal{Z}$ is a normalization constant that should be defined from the condition $2\pi \int_0^{\infty}dA A W_{\Sigma}^{(0)}=1$. Equation~(\ref{stationsolution}) shows the probability of the CNT to vibrate with an amplitude $A$ in the hybrid SC nanoelectromechanical device presented in Fig.~\ref{Fig1}. In the following, we will analyze the stationary solution of Eq.~(\ref{stationsolution}) in different regimes.

\section{Cooling and pumping regimes}\label{cooling}

As one can see from Eq.~(\ref{difeq}), the Wigner function $W_{\Sigma}^{(0)}(A)$ always has an extremum at $A=0$, and at such amplitudes when the effective damping coefficient $\tilde\gamma (A)$ induced by the nonlinear coupling with the electronic degree of freedom vanishes, $\tilde \gamma (A)=0$. First, we would like to check the stability of the solution Eq.~(\ref{stationsolution}) for $A=0$. This can be done by checking the sign of the effective inverse (dimensionless) temperature, $\beta_{eff}=(T_{eff}/\omega_0)^{-1}=\mathcal{Z}(d^2/dA^2)\log W_{\Sigma}^{(0)}(A)|_{A=0}$. Note that the sign of $\beta_{eff}$ is fully determined by the effective damping $\tilde \gamma(0)$ since the sign of the coefficient $\tilde D$ is fixed for any parameter. The probability density $W_{\Sigma}(A)$ has a maximum at $A=0$, i.e., the ground state with $A=0$ is stable, if $\beta_{eff}>0$ (or when $\tilde \gamma (0) > 0$), and it has a minimum, i.e. the ground state with $A=0$ is unstable, if $\beta_{eff}<0$ (or when $\tilde \gamma (0) < 0$). More precisely, the sign of $\beta_{eff}$ should be defined by the comparison of the effective damping $\tilde\gamma(0)$ with the real one $(\zeta^2 Q)^{-1}$, where $Q\sim 10^5\div 10^6$ is the quality factor of the CNT. However, we neglect the actual damping for simplicity below. 

We refer to the regime with $\beta_{eff}<0$ as a {\it pumping regime} and that with $\beta_{eff}>0$ as a {\it cooling regime}. Using Eqs.~(\ref{coefficients})--(\ref{coefficients1}) and (\ref{stationsolution}), it is easy to find the effective inverse temperature as follows:
\begin{widetext}
\begin{eqnarray}\label{beta}
\beta_{eff}= 8\kappa\frac{\omega_0\varepsilon_0\left[2\Delta_d^2(3\Gamma^2+\xi^2)-\Gamma^2(4\varepsilon_0^2+\Gamma^2)\right]}{(4\varepsilon_0^2+2\Delta_d^2+\Gamma^2)(\Gamma^4+4\varepsilon_0^2\Gamma^2+4\Delta_d^2\xi^2+8\Delta_d^2\Gamma^2)-16\Delta^2_d\varepsilon_0^2(\xi^2+3\Gamma^2)},
\end{eqnarray}
\end{widetext}
where $\xi=2\sqrt{\varepsilon_0^2+\Delta_d^2}$ is the Andreev energy level difference \cite{andreevlevel},\cite{andreevlevel2}. 
In Fig.~\ref{Fig2}, one can see a stability diagram (effective inverse temperature) as a function of different system parameters: the relative position of the electronic level in the QD and the strength of the tunnel coupling between the CNT and SC lead. The red and blue color schemes correspond to the pumping and cooling regimes, respectively.
Interestingly, the stability diagram may be obtained "quasiclassically" by considering the CNT's displacement and momentum as classical variables; see details in the Appendix. 

The bias voltage affects the stationary regime of the CNT's oscillations only through resolving the direction of the electronic flow since 
$\tilde \gamma(A)\propto\kappa$. Below, we explore the case of $V>0$ ($\kappa=+1$) only. The inverse effective temperature changes its sign when $\varepsilon_0=0$ and when the condition $2(4\varepsilon^2_0+4\Delta_d^2+3\Gamma^2)=(\Gamma/\Delta_d)^2(4\varepsilon^2_0+\Gamma^2)$ is fulfilled, see the numerator of Eq.~(\ref{beta}). More specifically, one has a transition between cooling and 
pumping at $\Gamma/\Delta_d\approx 2.37$ if $\varepsilon_0/\Delta_d\rightarrow 0$, and at $\Gamma/\Delta_d\approx \sqrt{2}$ if $\varepsilon_0/\Delta_d\rightarrow \infty$. The latter coincides with the instability condition in similar hybrid SC--normal metal NEMS, see Eq.~(13) in~ \cite{parafiloSC}.

\begin{figure}
\centering
\begin{tabular}{c}
\includegraphics[width=1\columnwidth]{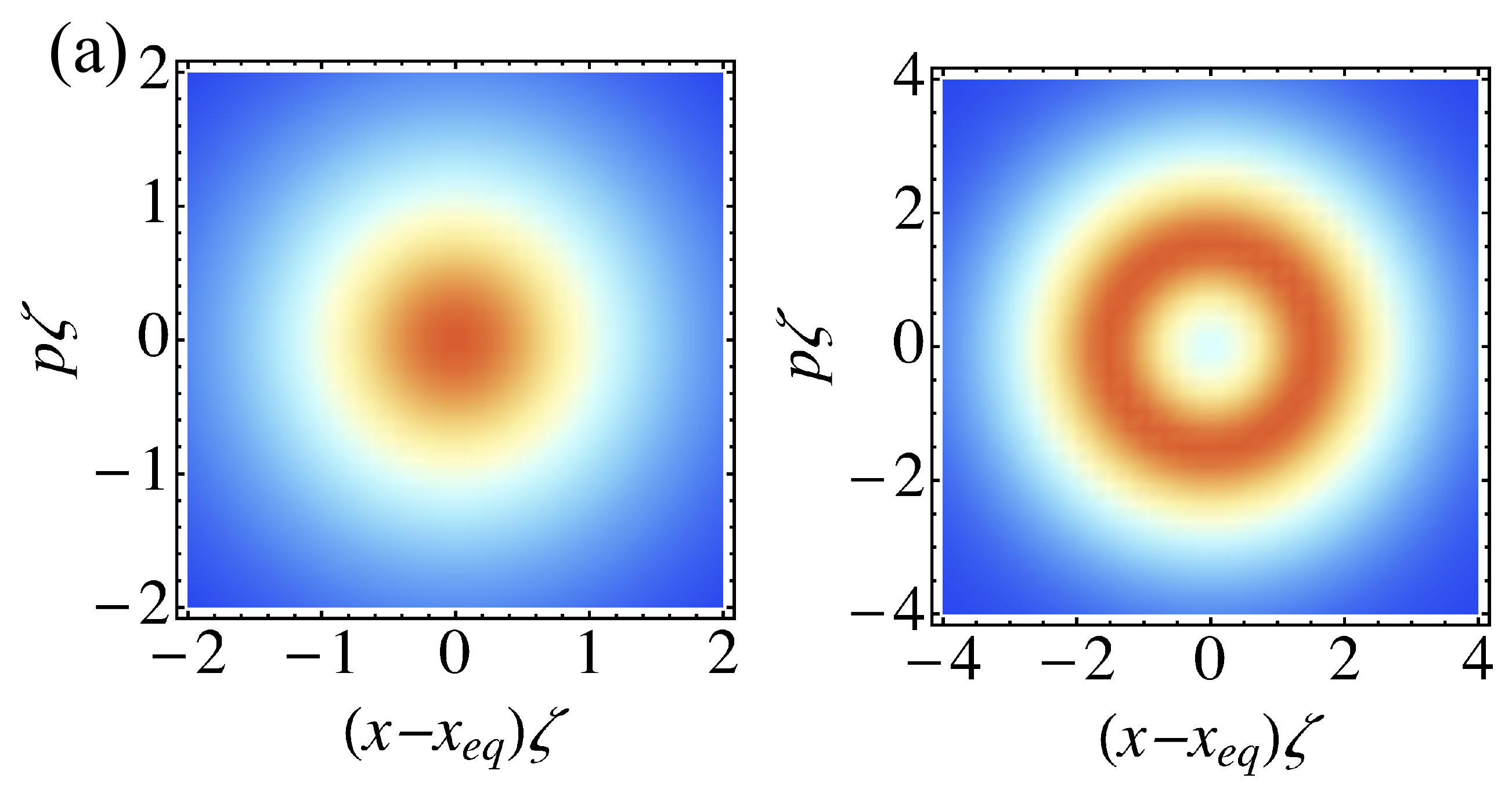}
\\
\includegraphics[width=1\columnwidth]{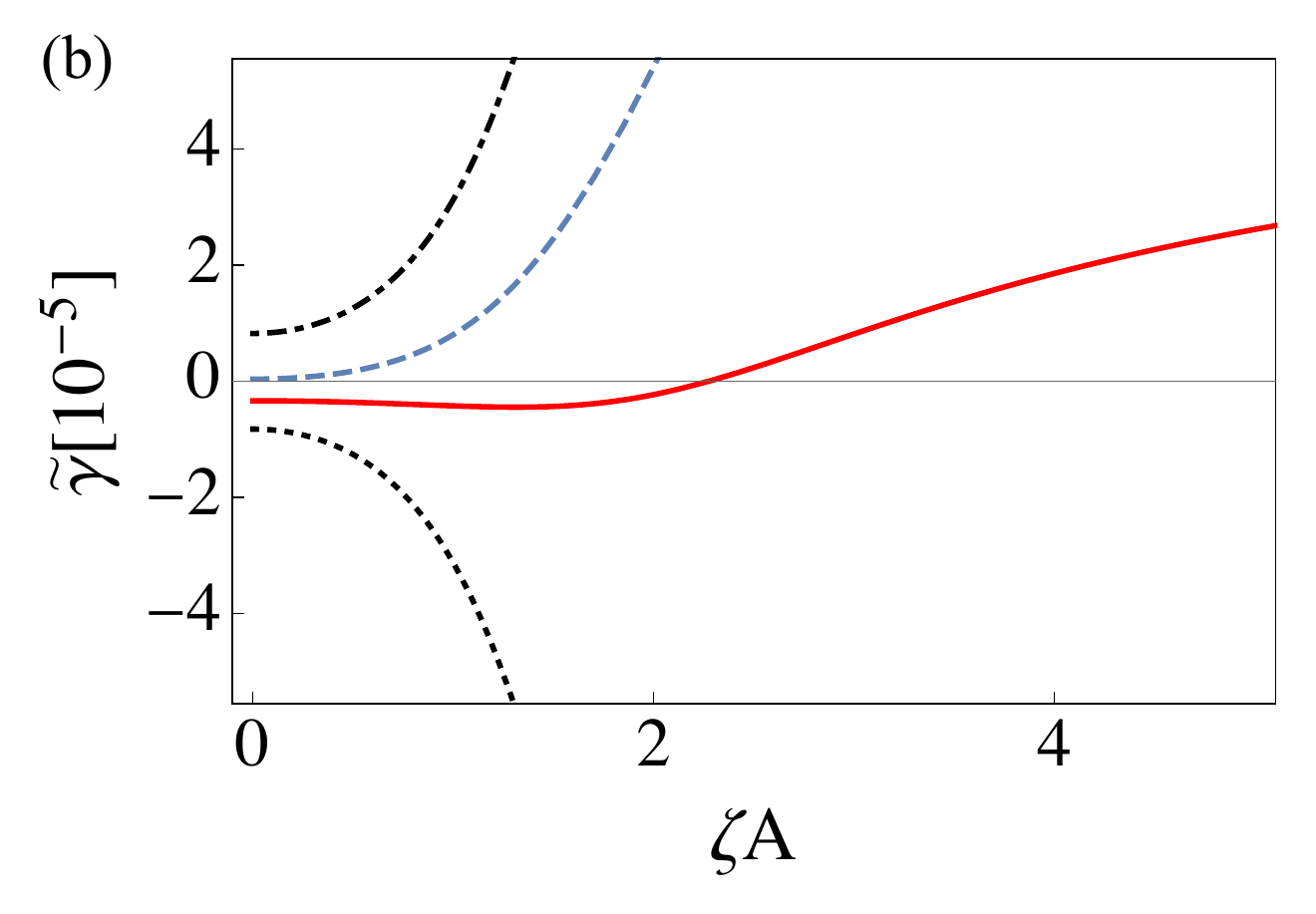}
\end{tabular}
\caption{(a) Examples of the Wigner function obtained using Eq.~(\ref{stationsolution}) as a function of dimensionless displacement and momentum for the cooling regime (left panel) and pumping regime (right panel). The maximum value of the Wigner function corresponds to the stable solution with $A=0$ in the cooling regime and to some finite $A_M$ in the pumping regime. (b) Values of the effective damping term $\tilde\gamma(A)$ as a function of the dimensionless amplitude $\zeta A$ for different sets of parameters: $\varepsilon_0/\Delta_d=4$, $\Gamma/\Delta_d=10$ (red solid line), $\varepsilon_0/\Delta_d=4$, $\Gamma/\Delta_d=\sqrt{2}$ (blue dashed line), $\varepsilon_0/\Delta_d=-4$, $\Gamma/\Delta_d=0.5$ (black dotted line), and $\varepsilon_0/\Delta_d=4$, $\Gamma/\Delta_d=0.5$ (black dash-dotted line). Only the case of $\kappa=+1$ and $\zeta=10^{-2}$ was considered.}
\label{Fig3}
\end{figure}

Next, we analyze the value of the minimum effective temperature in the cooling regime (blue regions in Fig.~\ref{Fig2}). This regime corresponds to a bell-shaped Wigner function with a maximum at $A=0$ and width determined by $\beta_{eff}$, as shown in the left panel of Fig.~\ref{Fig3}(a). We note that the effective damping coefficient $\tilde\gamma(A)$ is always a positively defined function, as one can see in Fig.~\ref{Fig3}(b) with the dashed and dash-dotted curves. In the adiabatic regime $\omega_0\ll \Gamma$, we obtain the effective temperature $T_{eff}$ using Eq.~(\ref{beta}): $T_{eff}\approx\Gamma /2$ if $\Gamma\gg \Delta_d$ (the minimum value of $T_{eff}$ is achieved at $\varepsilon_0=-\Gamma/2$), and $T_{eff}\approx \Gamma/\sqrt{2}$ if $\Gamma\ll \Delta_d$ (the minimum value of $T_{eff}$ is achieved at $\varepsilon_0=\sqrt{2}\Delta^2_d/\Gamma$). In the non-adiabatic regime, $T_{eff}$ may be obtained directly from Eqs.~(\ref{eq1})--(\ref{eq3}) using the perturbation theory for small amplitudes. At the most interesting case of resonance, when the frequency of the mechanical oscillations coincides with the Andreev energy level difference,  $\omega_0=2\sqrt{\varepsilon_0^2+\Delta_d}$, the effective temperature reads $T_{eff}=\Gamma/\sqrt{2}$ at $\varepsilon_0=\Gamma/2\sqrt{2}$ in the limit when $\Delta_d\ll \Gamma$.

The stationary state of Eq.~(\ref{stationsolution}) with $A=0$ in the pumping regime when $\beta_{eff}<0$  becomes unstable, and the CNT vibrations
develop into pronounced self-sustained oscillations of finite amplitude $A_M$. Interestingly, the appearance of the finite amplitude $A_M$ even in the absence of real damping $Q^{-1}=0$ is associated with the emergence of a new extremum of $W_{\Sigma}^{(0)}$ that corresponds to the condition $\tilde \gamma(A_M)=0$. We refer to such a phenomenon as a {\it self-saturation effect}. In this regime, the Wigner function is ring-shaped and has a maximum at $A_{M}$, as shown in the right panel of Fig.~\ref{Fig3}(a). 

\begin{figure}
\centering
\includegraphics[width=1\columnwidth]{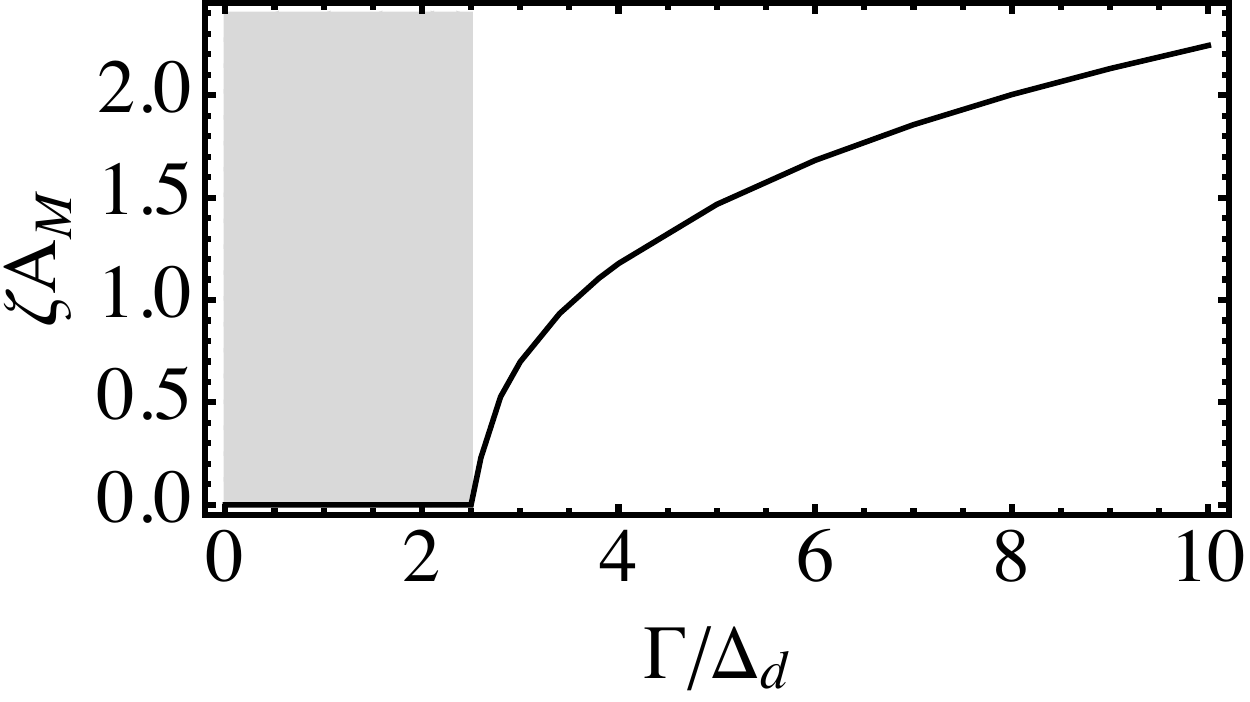}
\caption{Values of the self-saturated (maximum) amplitude of CNT oscillation as a function of normalized QD energy level width $\Gamma/\Delta_d$ for $\varepsilon_0/\Delta_d=0.5$. The gray region indicates the cooling regime of the mechanical subsystem.}
\label{Fig4}
\end{figure}

In Fig.~\ref{Fig3}(b), one can see the effective damping coefficient $\tilde \gamma (A)$ as a function of the CNT bending amplitude for different sets of system parameters. Here, one can recognize two different scenarios for reaching the self-saturation effect depending on the sign of the product of the QD energy level and direction of electronic flow, $\kappa \varepsilon_0$. The first scenario corresponds to the case when $\Delta_d\ll \Gamma$ and $\kappa \varepsilon_0 >0$, see the red solid line in Fig.~\ref{Fig3}(b). The CNT-QD vibrational ground state with $A=0$ is unstable (since $\tilde\gamma(0)<0$) and the amplitude of the CNT bending grows until the condition $\tilde\gamma(A)>0$ is fulfilled. The mechanical subsystem enters the cooling regime accompanied by CNT amplitude saturation until the condition $\tilde\gamma(A)<0$ is achieved, after which the mechanical subsystem enters the pumping regime again. As a consequence of such repetitions, a stationary regime with $\tilde\gamma(A_M)=0$ is established. Thus, the value of the self-saturated amplitude $A_M$ is determined by the parameters of the system. In Fig.~\ref{Fig4}, one can see values of the self-saturation amplitude as a function of the ratio between the tunnel coupling with the SC lead and the energy level width $\Gamma/\Delta_d$. The gray region in Fig.~\ref{Fig4} corresponds to the cooling regime. 

In the second scenario occurring when $\Delta_d\gg \Gamma$ and $\kappa \varepsilon_0<0$, see the black dotted line in Fig.~\ref{Fig3}(b), the mechanical subsystem with $A=0$ is unstable, while the development of CNT bending is not intrinsically limited. Thus, the saturation of the CNT amplitude is fully determined by the coupling with the thermodynamic environment, $Q\neq 0$.  

\section{Conclusions}\label{conclusion}

In conclusion, we investigated the stationary regimes of the mechanical subsystem in a hybrid nanoelectromechanical device comprising a carbon nanotube (CNT) suspended between two normal metal leads, equally biased by a voltage, and weakly coupled to a nearby superconducting (SC) STM tip. The proximity effect with the SC lead determined by the tunneling processes provides a coupling between the electronic and mechanical degrees of freedoms. As a consequence, the exchange of two electrons (Cooper pair) between the SC lead and the single-level quantum dot (QD) formed by the CNT results in the appearance of a specific force that acts on the CNT bending, which is similar to the force in the covalent bonds known in chemistry. 
Our analysis showed that the existence of two distinct cooling and pumping regimes of the mechanical subsystem is fully determined by the direction of the electronic flow, the relative position of the single electron energy level, and the ratio between the energy level width and the strength of the intra-QD electron pairing. The effective temperature of the mechanical subsystem was predicted in the regime of mechanical cooling.
This work demonstrated that peculiar quantum dynamics of the Cooper pair state forming in the QD in the pumping regime results in self-saturated bending oscillations of the CNT resonator.

\section*{Acknowledgement}
A.V.P. acknowledges the hospitality of the University of Gothenburg and Chalmers University of Technology. This work was supported by the Institute for Basic Science in the Republic of Korea (IBS-R024-D1) and Korean Institute for Advanced Study.

\begin{appendix}
\setcounter{equation}{0}
\setcounter{figure}{0}
\setcounter{table}{0}
\makeatletter
\renewcommand{\theequation}{S\arabic{equation}}
\renewcommand{\thefigure}{S\arabic{figure}}
\renewcommand{\bibnumfmt}[1]{[S#1]}
\renewcommand{\citenumfont}[1]{[S#1]}
\numberwithin{equation}{section}

\section{"Semiclassical" treatment}\label{App}
In this Appendix we derive Newton's equation for the CNT's displacement by using the reduced density matrix technique. Since we are interested in a classical consideration, we treat the operators of the CNT displacement $\hat x$ and momentum $\hat p$ as numbers after tracing them with the reduced density matrix: $x_c={\rm Tr}\{\hat x \hat \rho\}$, $p_c={\rm Tr}\{\hat p \hat \rho\}$. The system of the relevant equations for the reduced density matrices $R_1=\langle 0|\hat \rho |2\rangle+\langle 2|\hat \rho |0\rangle$, $R_2=i\langle 2|\hat \rho |0\rangle-i\langle 0|\hat \rho |2\rangle$, and $R_3=\langle 2|\hat \rho |2\rangle-\langle 0|\hat \rho |0\rangle$ read as follows:
\begin{eqnarray}\label{system1}
&&\dot R_1=-\Gamma R_1-2\varepsilon_0 R_2,\\
&&\dot R_2=2\varepsilon_0 R_1-\Gamma R_2+2\Delta_d e^{\zeta x_c}R_3,\\
&&\dot R_3 = -2 \Delta_d e^{\zeta x_c}R_2-\Gamma R_3+\kappa\Gamma.\label{system3}
\end{eqnarray}
Using the perturbation theory over small CNT displacement, the reduced density matrices can be expanded as $R_i(t)=R_i^{(0)}+R_i^{(1)}(t)$ ($i=1,2,3$). We find the stationary solution of the system in Eqs.~(\ref{system1})--(\ref{system3}) as
\begin{eqnarray}
R_{1}^{(0)}=-\kappa\frac{4\varepsilon_0\Delta_d}{D},R_{2}^{(0)}=\kappa\frac{2\Gamma\Delta_d}{D},R_{3}^{(0)}=\kappa\frac{4\varepsilon_0^2+\Gamma^2}{D},\nonumber\\
\end{eqnarray}
where $D=\xi^2+\Gamma^2$ and $\xi=2\sqrt{\varepsilon_0^2+\Delta_d^2}$ is the Andreev energy level distance. Introducing the vector $|{\bf R}^{(1)}\rangle=(R_1^{(1)}, R_2^{(1)}, R_3^{(1)})^T$, we obtain in the first order of perturbation theory:
\begin{eqnarray}\label{equation}
|\dot {\bf R}^{(1)}\rangle=-\hat m |{\bf R}^{(1)}\rangle+2\zeta\Delta_dx_c(t)|e\rangle,
\end{eqnarray}
where $|e\rangle=(0, R_3^{(0)}, -R_2^{(0)})^T$ and
\begin{eqnarray}
\hat m =\begin{pmatrix}
\Gamma & 2\varepsilon_0 & 0 \\-2\varepsilon_0 & \Gamma & -2\Delta_d \\
0 & 2\Delta_d & \Gamma
\end{pmatrix}.
\end{eqnarray}
The solution of Eq.~(\ref{equation}) can written as follows:
\begin{eqnarray}\label{solut}
|{\bf R}^{(1)}\rangle=2\zeta\Delta_d \int_0^{\infty}d\tau [x_c(t)-\tau \dot x_c (t)]\hat U e^{-\hat E \tau}\hat U^{-1}|e\rangle,\nonumber\\
\end{eqnarray}
where
\begin{eqnarray}
\hat U =\begin{pmatrix}
-\Delta_d & 2\varepsilon_0 & 2\varepsilon_0 \\ 0 & i\xi & -i\xi \\
\varepsilon_0 & 2\Delta_d & 2\Delta_d
\end{pmatrix},
\end{eqnarray}
and $\hat E$ is a matrix with the eigenvalues of the matrix $\hat m$ ($\lambda_1=\Gamma$, $\lambda_{2,3}=\Gamma \pm i \xi$) as diagonal elements. Here, in the adiabatic regime $\omega_0\ll \Gamma$, a small retardation effect of the CNT displacement $x_c(t-\tau)\approx x_c(t)-\tau \dot x_c(t)$ is taken into account.
As a result, after taking the integral in Eq.~(\ref{solut}) we obtain 
\begin{eqnarray}\label{solut2}
&&R^{(1)}_1\approx\dot x_c (t)2\zeta\kappa\Delta_d\varepsilon_0\frac{\{4\Gamma^2(\Gamma^2+4\varepsilon^2_0)-8\Delta_d^2(\xi^2+3\Gamma^2)\}}{\Gamma D^3}.\nonumber\\
\end{eqnarray}

The equations of motion for the CNT's displacement and momentum read as follows:
\begin{eqnarray}
&&\dot x_c =\omega_0 p_c,\\
&&\dot p_c=-\omega_0 x_c + \zeta \Delta_d e^{\zeta x_c} R_1(t).
\end{eqnarray}
Combining them with the result obtained above in Eq.~(\ref{solut2}), we find
\begin{widetext}
\begin{eqnarray}\label{FPequation2}
\ddot x_c +\omega_0 \zeta^2\kappa 2\Delta_d^2 \varepsilon_0\frac{\{8\Delta_d^2(\xi^2+3\Gamma^2)-4\Gamma^2(\Gamma^2+4\varepsilon_0^2)\}}{\Gamma D^3}\dot x_c +\tilde \omega^2_0 x_c =-\zeta\kappa\omega_0\Delta_d^2\frac{4\varepsilon_0}{D},
\end{eqnarray}
\end{widetext}
where $\tilde \omega_0$ is the interaction renormalized frequency of the mechanical oscillations. The effective damping induced by interaction with the electronic degree of freedom coincides with the coefficient $\tilde \gamma(0)$ from the Eq.~(\ref{stationsolution}). The imaginary part of the eigenfrequencies of Eq.~(\ref{FPequation2}) is negative at $\Delta_d\ll \Gamma$ and $\kappa\varepsilon_0>0$, which corresponds to the existence of instability in the mechanical subsystem. This regime corresponds to the pumping regime discussed in the main text. On the other hand, the positive imaginary part of the eigenfrequencies occurring at $\Delta_d\gg \Gamma$ and $\kappa \varepsilon_0>0$ corresponds to the damping regime, which is connected with the cooling regime in the main text.

\end{appendix}

\vspace*{3mm}


\end{document}